\documentclass[runningheads]{llncs}
\usepackage{caption}
\usepackage{lipsum,setspace,gensymb,graphicx,array,amsmath,caption,epstopdf,amsfonts,algorithm,algorithmic}
\usepackage{subfigure,psfrag,epsfig,amssymb,array,flushend}
\begin{document}
\fontsize{10pt}{10pt}\selectfont
%\lipsum[1]
%
\title{Blind Robust 3-D Mesh Watermarking \\ based on Mesh Saliency and QIM quantization \\ for Copyright Protection}
\author{Mohamed Hamidi\inst{1}%\orcidID{0000-0001-9758-3131} 
\and
Aladine Chetouani\inst{2} \and
Mohamed El Haziti\inst{1} \and Mohammed El Hassouni\inst{1}  \and Hocine Cherifi\inst{3}}
\authorrunning{Mohamed Hamidi et al.}
\institute{LRIT - CNRST URAC29, Rabat IT Center, Faculty of Sciences, Mohammed V University in Rabat, Morocco
\email{\{hamidi.medinfo,elhazitim\}@gmail.com}
 \and
PRISME Laboratory, University of Orleans, France
\email{aladine.chetouani@univ-orleans.fr}\\
 \and
LRIT - CNRST URAC29, Rabat IT Center, FLSH ,\\ Mohammed V University in Rabat, Morocco
\email{mohamed.elhassouni@gmail.com} 
 \and
 Le2i, UMR 6306 CNRS, University of Burgundy, France
\email{hocine.cherifi@u-bourgogne.fr}}
\maketitle              % typeset the header of the contribution
\begin{abstract}
Due to the recent demand of 3-D models in several applications like medical imaging, video games, among others,  the necessity of implementing 3-D  mesh watermarking schemes  aiming to protect copyright has increased considerably. The majority of robust 3-D watermarking techniques have essentially focused on the robustness against attacks while the  imperceptibility of these techniques is still a real issue. In this context, a blind robust 3-D mesh watermarking method based on mesh saliency and Quantization Index Modulation (QIM) for Copyright protection is proposed. 
%The watermarking primitives of the proposed scheme are the vertex norms. 
The watermark is embedded by quantifying the vertex norms of the 3-D mesh using QIM scheme since it offers a good robustness-capacity tradeoff. The choice of the  vertices is adjusted by the mesh saliency to achieve watermark robustness and to avoid visual distortions. The experimental results show the  high imperceptibility of the proposed scheme while ensuring a good robustness  against a wide range of attacks including additive noise, similarity transformations, smoothing, quantization, etc.
\keywords{3-D mesh watermarking  \and Quantization Index Modulation(QIM) \and Copyright protection \and mesh saliency.}
\end{abstract}
\section{Introduction}
Due to the rapid development of digital  services and the increase in network bandwidth, the transfer of multimedia contents such as image, audio, video and 3-D model has been increased considerably. These contents can be modified or duplicated  easily. Therefore, the need to develop security methods became crucially important. Digital watermarking has been found as an efficient solution to overcome this issue. Its underlying concept is to embed an extra information called watermark  into multimedia content to protect its ownership.  
	In the last decade, 3-D meshes have been widely used in medical images, computer aided design (CAD), video games, virtual reality, etc. 
%	A mesh can be defined as a collection of polygonal facets that aim to approximate a real 3-D object. 
	Each 3-D watermarking system should ensure a three major requirements : imperceptibility, capacity and robustness. 
%	Imperceptibility refers to the similarity between the original 3-D mesh and the watermarked one. Capacity means the maximum amount of information that can be inserted in the 3-D model, while robustness represents the ability of extracting the watermark even if the watermarked 3-D model has incurred changes called attacks. 
	The attacks  can be divided into two major types. Connectivity attacks  including subdivision, cropping, remeshing,  and simplification. Geometric attacks that include similarity transformations, local deformation operations and signal processing manipulations. The applications of 3-D mesh watermarking include copyright protection, authentication, content enhancement, indexation, etc. We note that the proposed method aims to protect copyright. It is worth mentioning that in contrast with the maturity of image watermarking techniques \cite{hamidi2018hybrid}, there are only few watermarking methods that work on 3-D meshes. In addition, the processing techniques applied to 2D cannot be used in case of 3D meshes \cite{eude1994statistical}. This is due to the challenges in three dimensional geometry related to its irregular topology as well as the complexity of attacks that target this kind of geometrical content \cite{Survey2007}. The majority of 3-D watermarking techniques have essentially focused on the robustness against attacks.
Few 3-D mesh watermarking methods based on saliency have been proposed \cite{hamidi2019blind}. In \cite{nakazawa2010visually}, a watermarking 3-D mesh method using the visual saliency is presented. Firstly, the perceptually conspicuous regions using the mesh saliency \cite{lee2005mesh} have been identified. Secondly, the norm of each vertex is calculated and its histogram is constructed. The watermark is embedded in each bin by normalizing the associated vertex norms. Zhan et al. \cite{zhan2014blind} proposed a blind 3-D mesh watermarking algorithm based on curvature. The authors calculated the root mean square curvature for all vertices of the 3-D model. 
The watermark is embedded by modulating the mean of the root mean square curvature fluctuation of vertices. Rolland-Neviere et al. \cite{rolland2014triangle} proposed a 3-D mesh watermarking method where the watermark embedding is formulated as a quadratic programming problem.  
Jeongho Son et al. \cite{son2017perceptual} proposed a 3-D watermarking method with the aim of preserving the appearance of the watermarked 3-D model. The  method used  the distribution of the vertex norm histogram as a watermarking primitive  that  
is already introduced by Cho et al. \cite{Cho2007}. The latter inserts the watermark by altering the mean or variance of the vertex norms histogram. 

In this paper, a 3-D mesh  blind and robust watermarking method based on mesh saliency and QIM quantization is proposed. The watermark bits are inserted in the host 3-D mesh by quantizing its vertices norms. The choice of these norms has been guided by the mesh saliency of the 3-D mesh. Taking the full advantages of QIM scheme as well as mesh saliency, the proposed method can achieve high robustness to common attacks while preserving high imperceptibility. The rest of this paper is organized as follows. Section \ref{Background} presents the background. Section \ref{Proposed method} gives a description of the proposed method composed by embedding and extraction. The experimental setup, evaluation metrics and experimental results are discussed in Section \ref{Experimental results}. Finally, Section \ref{Conclusion} concludes the paper.

\section{Background}
\label{Background}
\subsection{3-D mesh saliency}
%Nowadays, saliency detection becomes an interdisciplinary scientific study of computer science and  human perception. 
%It enables to automatically  detect perceptually important points or regions of a 3-D mesh \cite{song2014mesh}. 
Mesh saliency can be defined as a  measure that captures the importance of a point or local region of a 3-D mesh in a similar way to human visual perception\cite{song2014mesh}. 
% This technique generally merges perceptual criteria inspired by  human visual system (HVS) with mathematical measures based on geometry.
  The visual attention of Human is usually directed to the salient shape of the 3-D model. The evaluation of mesh saliency used  in the proposed scheme is Lee et al. \cite{lee2005mesh}. The later evaluates the saliency of each vertex using the difference in mean curvature of the 3-D mesh surfaces from those at other vertices  in the neighborhood. The first step is computing surface curvatures. The computation of the curvature at each vertex $v$  is performed using Taubin's method \cite{taubin1995estimating}. Let $Curv(v)$ the mean curvature of a mesh at a vertex $v$. The Gaussian-weighted average of the mean curvature can be expressed as follows:  
\begin{equation}
G(Curv(v),\sigma )=\frac{\sum_{x \in N(v,2\sigma)} Curv(x)exp(\frac{-\parallel x-v \parallel^2}{2\sigma^2})}{ \sum_{x \in N(v,2\sigma)}exp(\frac{-\parallel x-v \parallel^2}{2\sigma^2})}
\label{eq:GaussAvergMeanCurvature}
\end{equation}
where $x$ is a mesh point and $N(v,\sigma)$ denotes the neighborhood for a vertex $v$ which represents a set of points within an Euclidean distance $\sigma$ calculated as : 
\begin{equation}
N(v,\sigma)=\{x | \parallel x-v \parallel < \sigma\}
\end{equation}
The saliency $S(v)$ of a vertex $v$ is calculated as the absolute difference between the Gaussian-weighted averages computed at fine and coarse scale. 
\begin{equation}
S(v)=|G(Curv(v),\sigma) - G(Curv(v),2\sigma)|
\end{equation}

%Saliency of 3D test meshes
\begin{figure}[h!]
    \centering
    \subfigure[]{\label{sub1} \includegraphics[width=0.10\textwidth]{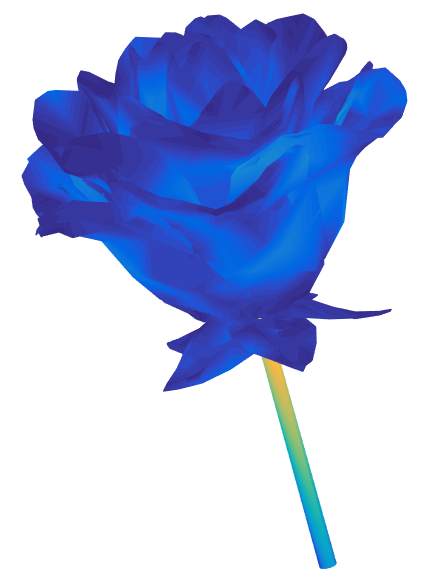}}
    \subfigure[]{\label{sub2} \includegraphics[width=0.10\textwidth]{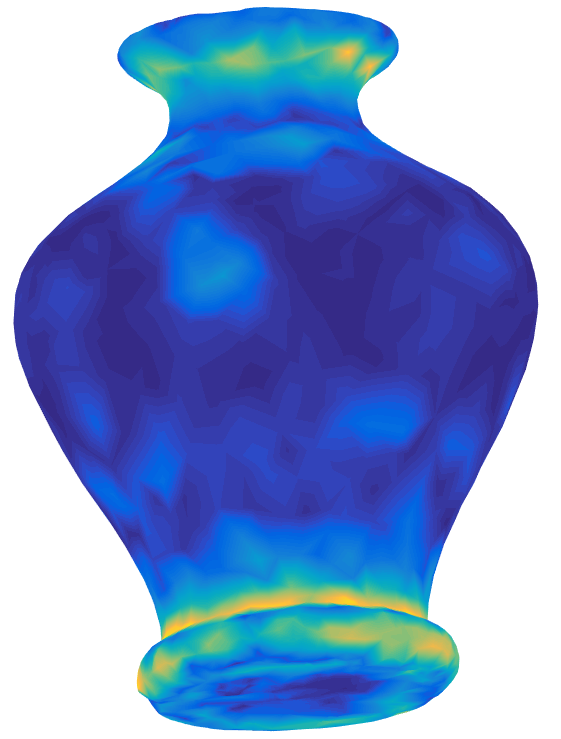}}
    \subfigure[]{\label{sub3} \includegraphics[width=0.10\textwidth]{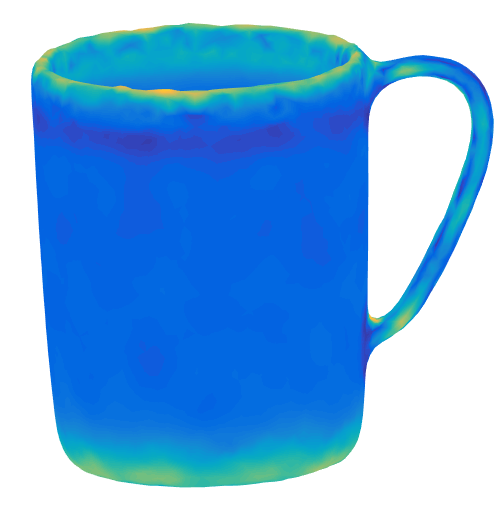}}
%    \subfigure[]{\label{sub4} \includegraphics[width=0.09\textwidth]{Objects_saliency/antS.png}}
    \subfigure[]{\label{sub5} \includegraphics[width=0.10\textwidth]{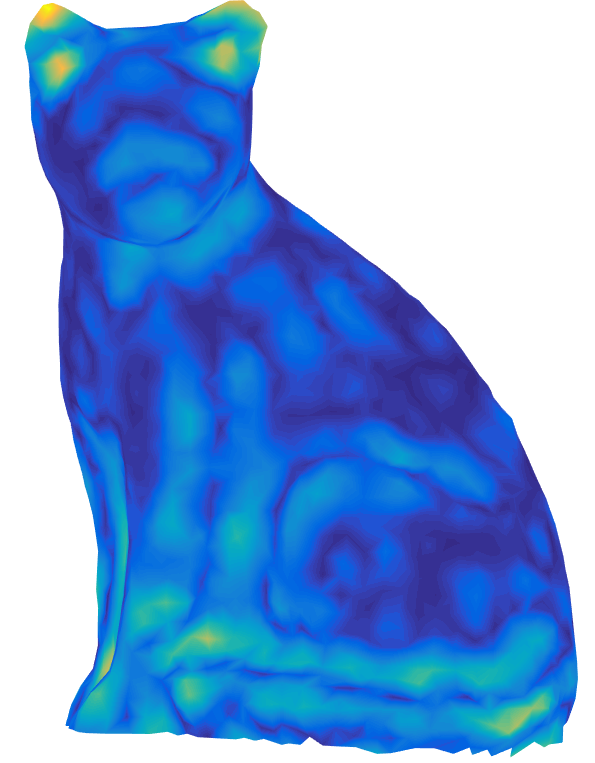}}
%     \subfigure[]{\label{sub6} \includegraphics[width=0.09\textwidth]{Objects_saliency/bunnyS.png}}
       \caption{ Saliency of 3-D meshes using Lee's method \cite{lee2005mesh} (a) Flower, (b) Vase, (c) Cup, (d) Cat.}
    \label{fig:SaliencyObjects}
\end{figure}
Figure \ref{fig:SaliencyObjects} shows an example of mesh saliency using Lee's method \cite{lee2005mesh}.
%\begin{figure}[!h]
%\centering
%\captionsetup{justification=centering}
%%\begin{center}
%\includegraphics[width=0.25\textwidth]{images/SaliencyVsMesh.pdf}
%%\end{center}
%\caption{3-D meshes and their corresponding mesh saliency : (left) Original 3-D meshes, (right) 3-D mesh saliency.}
%\label{fig:SaliencyVsMeshes}
%\end{figure}
%
\subsection{Quantization Index Modulation}
Quantization Index Modulation (QIM) approaches have been widely used  in image, audio , and video processing. Their application to 3-D meshes is trivial since two quantifiers are needed to insert a binary message in the 3-D data \cite{chen2001quantization}.
 QIM techniques are simple to implement and have a small complexity. In addition, they ensure a high tradeoff between robustness and capacity. Each watermark bit is associated to a quantizer in the host signal. Let $b\in\left( 0,1 \right)$ the watermark bit and $x$ the host signal to be quantized. The QIM techniques operate independently on these two elements. In order to embed a bit $b$, two quantizers $Q_0$ and $Q_1$ are needed \cite{vasic2013simplification}. They can be defined as follows: 
\begin{equation}
Q_b(x)=\Delta[\frac{1}{\Delta}(x-(-1)^b \frac{\Delta}{4})+(-1)^b \frac{\Delta}{4}]
\label{eq:Quantizers}
\end{equation}
Where $[]$ refers to the rounding operation and $\Delta$ is the quantization step. 
%The later is crucial since it influences on the fidelity of the watermark embedding. In fact, optimal quantization step is still a challenging task in QIM approaches. 
%Indeed since the distortion decreases according to the quantization step.
% Indeed, when the quantization step is small the distortion in the mesh geometry decrease and vice versa. 
In the extraction process the signal is re-quantized using the same family of quantizer to get the embedded bits. The recovered bits are easily  calculated as follows: 
\begin{equation}
\hat{b}= arg min \left \| x-Q_b(x) \right \|
\label{eq:RecovredBits}
\end{equation}

\section{The proposed method}
\label{Proposed method}
\begin{figure}[!h]
%\centering
\subfigure[]{
\includegraphics[height =6.0cm]{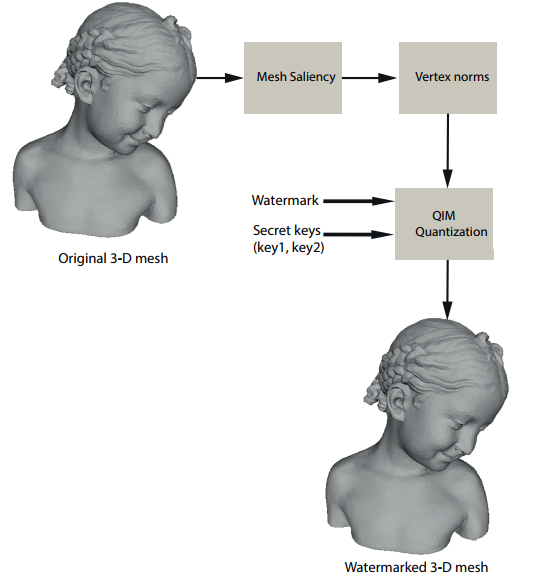}
}
\quad
\subfigure[]{
\includegraphics[height =4.0cm]{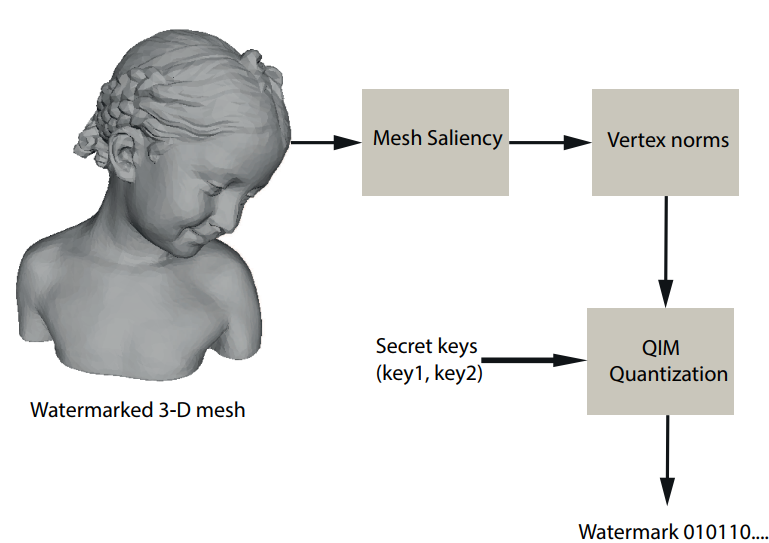}
}
\caption {Flowchart of the proposed method: (a) watermark embedding, (b) watermark extraction.}
\label{fig:FlowchartOurMethod}
\end{figure}

\begin{figure}[!h]
\centering
\captionsetup{justification=centering}
%\begin{center}
\includegraphics[width=0.55\textwidth]{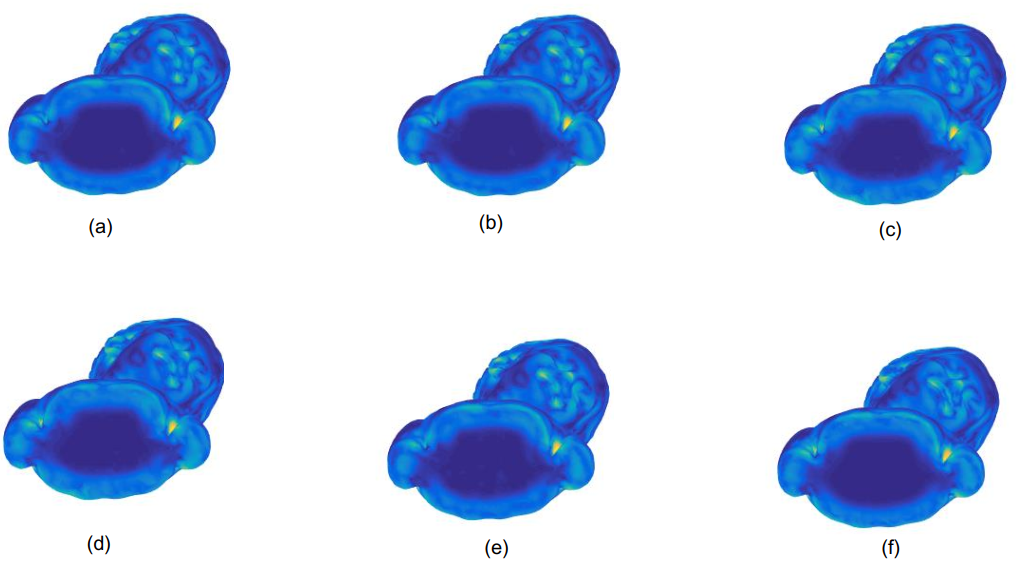}
%\end{center}
\caption{Mesh saliency of Bimba before and after attacks : (a)  before attack, (b) additive noise $0.3$\%, (c) Similarity transformation $1$, (d) Simplification 10\%, (e) Quantization $9$ bits, (f) Smoothing $\lambda=0.1$  ($30$ iterations).}
\label{fig:StabilitySaliencyAttacks}
\end{figure}

In this paper, a blind robust 3-D mesh watermarking technique based on visual saliency and QIM quantization for Copyright protection is proposed. 
%The mesh geometry is a good candidate for watermark embedding since it does not affect the overall shape of the model with respect to small modifications in the local geometry and ensures good robustness against a wide range of attacks including smoothing, translation, rotation, scaling, etc.
 The watermark is embedded  by modifying the salient vertex norms of the 3-D mesh to using mesh saliency and QIM quantization. The mesh saliency of Lee et al. \cite{lee2005mesh} used to obtain candidate vertices aims to ensure high imperceptibility and to improve the watermark robustness. The choice of these points has been driven by the fact that these primitives are  relatively stable even after applying several attacks including similarity transformations, noise addition, smoothing, quantization, etc. Figure \ref{fig:FlowchartOurMethod} shows the flowchart of the proposed  method. Figure \ref{fig:StabilitySaliencyAttacks} illustrates the  Lee's \cite{lee2005mesh} mesh saliency  of Bimba model after applying different attacks. 
%The QIM quantization is utilized in our work since it offers a good capacity-robustness tradeoff \cite{chen2001quantization}. 
\subsection{Watermark embedding}
The proposed method takes the full advantage of the mesh saliency to achieve  high imperceptibility. The watermarking bits are embedded by quantizying the vertex norms of the 3-D model using mesh saliency. The motivation behind using the QIM quantization is the good tradeoff between capacity and robustness \cite{chen2001quantization}. In addition, QIM methods are blind. Firstly, the mesh saliency is computed and a threshold is fixed automatically to define the salient and non-salient points. In fact, for each saliency vector, the $70$\% maximum values represent the salient points while the other points are considered non-salient. Next, the norms of salient points are calculated according to this threshold. After, a watermark is generated using pseudo-random generator using a secret key (key1). Afterwards, two quantizers $Q_{zero}$ and $Q_{one}$ are calculated using equation \ref{eq:Quantizers} according to the watermark bit. Finally, starting from the modified vertex norms the new vertex coordinates are calculated using equation \ref{eq:MeshReconst} in order to construct  the 3-D watermarked mesh. We note that the quantization step can be considered as second secret key (key2) that will be used  in the extraction step.
\begin{equation}
{V}'({x}',{y}',{z}')=\frac{\left \| {V}' \right \|}{\left \| V \right \|} V(x,y,z)
\label{eq:MeshReconst}
\end{equation}

%\begin{figure}[!h]
%\centering
%\captionsetup{justification=centering}
%\begin{center}
%\includegraphics[width=0.26\textwidth]{images/Embedding_Scheme.png}
%\end{center}
%\caption{The proposed embedding scheme.}
%\label{fig:EmbeddingScheme}
%\end{figure} 

\subsection{Watermark extraction}
The watermark extraction is blind since only secret keys (key1 and key2) are needed. First, the mesh saliency of the 3-D watermarked model is calculated and the salient points are extracted according to the threshold used in the watermark embedding. We note that this parameter is chosen automatically since it represents the $70$\% maximum values of the saliency vector. Next, the norms according to the chosen vertices are calculated. The two quantizers are calculated and the extracted watermark bits are obtained using equation \ref{eq:RecovredBits}.

%\begin{figure}[!h]
%\centering
%\captionsetup{justification=centering}
%%\begin{center}
%\includegraphics[width=0.25\textwidth]{images/Extracting_Scheme.png}
%%\end{center}
%\caption{The proposed extracting scheme.}
%\label{fig:ExtractingScheme}
%\end{figure}

\section{Experimental results}
\label{Experimental results}
\subsection{Experimental setup}
\begin{figure}[h]
    \centering
    \subfigure[]{\label{sub21} \includegraphics[width=0.10\textwidth]{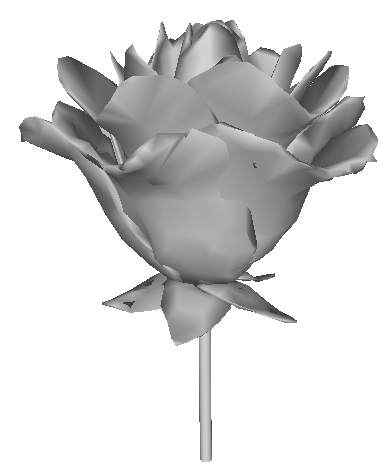}}
    \subfigure[]{\label{sub22} \includegraphics[width=0.10\textwidth]{images/flower.png}}
    \subfigure[]{\label{sub23} \includegraphics[width=0.09\textwidth]{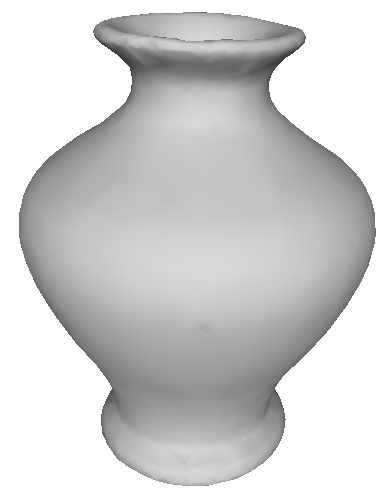}}
    \subfigure[]{\label{sub24} \includegraphics[width=0.09\textwidth]{images/vase.png}}
    \subfigure[]{\label{sub25} \includegraphics[width=0.10\textwidth]{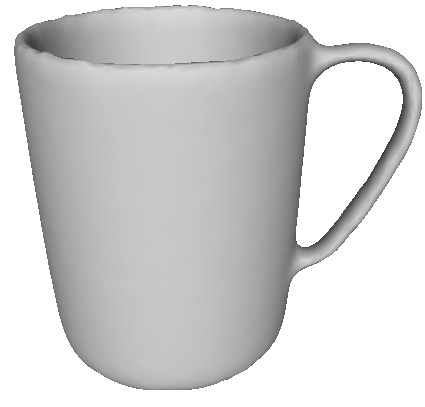}}
    \subfigure[]{\label{sub26} \includegraphics[width=0.10\textwidth]{images/cup.png}}
%    \subfigure[]{\label{sub27} \includegraphics[width=0.15\textwidth]{images/ant.png}}
%    \subfigure[]{\label{sub28} \includegraphics[width=0.15\textwidth]{images/ant.png}}
%    \subfigure[]{\label{sub29} \includegraphics[width=0.07\textwidth]{images/cat.png}}
%    \subfigure[]{\label{sub30} \includegraphics[width=0.07\textwidth]{images/cat.png}}
       \caption{(a) Flower, (b)Watermarked Flower, (c) Vase, (d) Watermarked Vase, (e) Cup, (f) Watermarked Cup.}
    \label{fig:OriginalAndWatermarkedObjects}
\end{figure}

The proposed watermarking method is tested on several 3-D meshes with different shape complexities: Flower ($2523$ vertices, $4895$ faces), Vase ($2527$ vertices, $5004$ faces), Cup ($9076$ vertices, $18152$ faces), Ant ($7654$ vertices, $15304$ faces), Bimba ($8857$ vertices,  $17710$ faces) and cat ($3534$ vertices, $6975$ faces) where some of them are shown in Fig. \ref{fig:OriginalAndWatermarkedObjects}((a),(c),(e)). It is  worth noticing that for comparison purpose the imperceptibility and robustness evaluation have been performed using the 3-D models: Bunny ($34835$ vertices, $69666$ faces), Horse ($112642$ vertices ,$225280$ faces) and Venus ($100759$ vertices, $201514$ faces). The quantization step is chosen is such a way that ensures good tradeoff between imperceptibility and robustness. This parameter is tuned experimentally and we kept $\Delta=0.08$. Several  metrics have been used to measure the amount of distortion introduced by the embedding process. This distortion can be measured geometrically or perceptually. The maximum root mean square error (MRMS) proposed in \cite{Metro} is used to calculate the objective distortion between the original meshes and the watermarked ones. The mesh structural distortion measure (MSDM) metric is chosen  to measure the visual degradation of the watermarked meshes \cite{lavoue2006perceptually}. 
%The MSDM value is equal to $0$ when the original and watermarked 3-D objects are identical . Otherwise, the MSDM value is equal to $1$ when the objects are visually very different.
 The robustness is measured using the normalized correlation ($Corr$) between the inserted watermark and the extracted one.

\begin{table}[!h]
\footnotesize
\centering
{\renewcommand{\arraystretch}{0.4}
%\captionsetup{justification=centering}
\caption{Watermark imperceptibility measured in terms of MRMS, HD and MSDM.}
\label{tab:imperceptibilityEvaluation}
%Résultats pour k=750
\begin{tabular}{|c|c|c|c|c|}
\hline
Model &  \: MRMS ($10^{-3}$) \: & \: HD ($10^{-3}$) \: & \: MSDM   \:  \tabularnewline
 \hline
Flower & $0.63$ & $4.33$ & $0.30$ \tabularnewline
\hline
Vase &$0.41$ & $3.34$ &$0.37$ \tabularnewline
\hline
Cup  & $0.98$&  $2.95$ & $0.37$\tabularnewline
\hline
Ant   & $0.62$ &$4.19$ &$0.51$ \tabularnewline
\hline

 Cat   & $0.61$ & $1.0$ &$0.16$\tabularnewline
\hline
Bimba   & $0.39$ & $1.63$ &$0.10$\tabularnewline
\hline

\end{tabular}}
\\
\end{table}

%\subsubsection{Robustness}
%~\\
%The robustness is measured using the normalized correlation ($Corr$) between the inserted watermark and the extracted one as given by the following equation : 
%\begin{equation}
%Corr=\frac {\sum_{i=1}^{M}(w'_i-\overline{w}^*)(w_i-\overline{w})}{\sqrt{\sum_{i=1}^{M}(w'_i-\overline{w}^*)^2.\sum_{i=1}^{M}(w_i-\overline{w})^2}}
%\end{equation}
%Where $i\in \left \{ 1,2,\dots, M \right \}$, $\overline{w}^*$ and $\overline{w}$ are the averages of the watermark bits respectively.

% It can be seen from this- figure that no perceptible distortions can be remarked in the watermarked 3D meshes.

\subsection{Results discussion}
\subsubsection{Imperceptibility}
  Fig. \ref{fig:OriginalAndWatermarkedObjects} illustrates the original and watermarked 3-D meshes. We can see that the distortion is very imperceptible. This is due to the saliency adjustment. In addition, according to Table \ref{tab:imperceptibilityEvaluation}, it can be observed that the proposed method can achieve high imperceptibility in terms of MRMS, HD and MSDM. We believe that this performance is obtained thanks to the exploitation of mesh saliency to avoid serious distortions. It can be also observed that the imperceptibility results in terms of MRMS, HD and MSDM are different from a mesh to another. This difference is mainly due to the curvature nature of each one of these 3-D meshes.

\begin{table}[!h]
\footnotesize
\centering
{\renewcommand{\arraystretch}{0.4}
%\captionsetup{justification=centering}
\caption{Watermark imperceptibility without using saliency measured in terms of MRMS, HD and MSDM  compared to the proposed method.}
\label{tab:ImperceptWithoutandWithSaliency}
%Résultats pour k=750
\begin{tabular}{|c|c|c|c|c|}
\hline
Model &  \: MRMS ($10^{-3}$) \: & \: HD ($10^{-3}$) \: & \: MSDM   \:  \tabularnewline
 \hline
Flower & $0.89/0.63$ & $5.03/4.33$ & $0.88/0.30$ \tabularnewline
\hline
Vase &$0.58/0.41$ & $4.76/3.34$ &$0.76/0.37$ \tabularnewline
\hline
Cup  & $1.02/0.98$&  $3.45/2.95$ & $0.87/0.37$\tabularnewline
\hline
Ant   & $0.83/0.62$ &$4.43/4.19$ &$1.0/0.51$ \tabularnewline
\hline

 Cat   & $1.2/0.61$ & $1.9/1.0$ &$0.29/0.16$\tabularnewline
\hline
Bimba   & $0.76/0.39$ & $2.98/1.63$ &$1.66/0.10$\tabularnewline
\hline

\end{tabular}}
\\
\end{table}

  To further evaluate the importance of using mesh saliency to improve the imperceptibility of the proposed method, we compare the obtained results with those obtained without using the saliency. Table \ref{tab:ImperceptWithoutandWithSaliency} exhibits the imperceptibility performance in terms of MRMS, HD and MSDM without using the mesh saliency compared to the proposed method based on mesh saliency. According to Table \ref{tab:ImperceptWithoutandWithSaliency}, it can be seen that the proposed method outperforms which illustrates the  imperceptibility improvement achieved using the saliency aspect in the watermark embedding.
\subsubsection{Robustness}
\begin{figure}[H]
 \centering
%    \subfigure[]{\label{sub13} \includegraphics[width=0.10\textwidth]{Attacked_Bimba/BimbOrig.png}}
    \subfigure[]{\label{sub14} \includegraphics[width=0.11\textwidth]{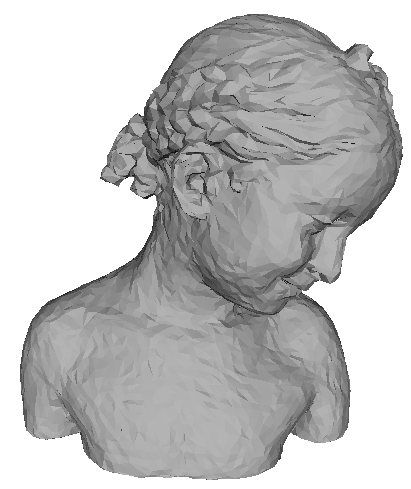}}
    \subfigure[]{\label{sub15} \includegraphics[width=0.11\textwidth]{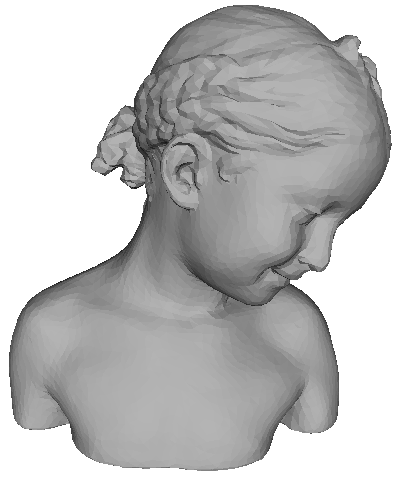}}
    \subfigure[]{\label{sub16} \includegraphics[width=0.10\textwidth]{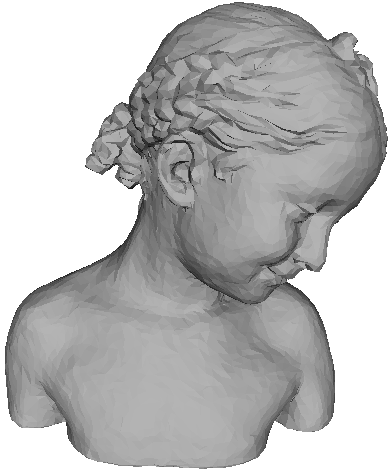}}
%    \subfigure[]{\label{sub17} \includegraphics[width=6.4cm]{Attacked_Bimba/Bimba_Simplification_1000.png}}
    \subfigure[]{\label{sub18} \includegraphics[width=0.13\textwidth]{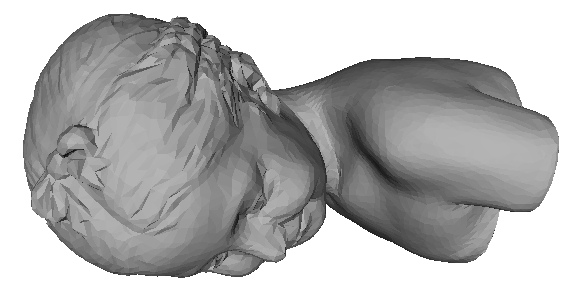}}
%     \subfigure[]{\label{sub19} \includegraphics[width=6.4cm]{Attacked_Bimba/Bimba_Subdiv_Midpoint00.png}}
     \subfigure[]{\label{sub19} \includegraphics[width=0.10\textwidth]{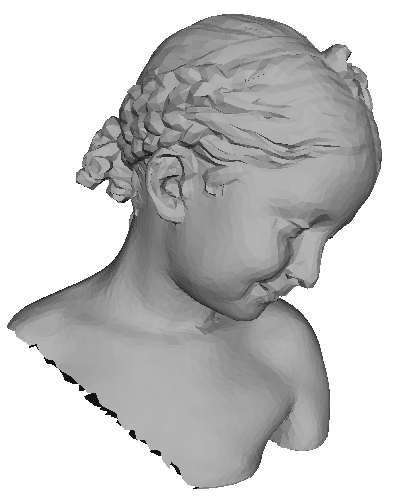}}
%     \subfigure[]{\label{sub7} \includegraphics[width=13.4cm]{Attacked_Bimba/Cropping_bimba.png}}
       \caption{Six attacks versions of Bimba: (a) noise addition $0.50$\%, (b) Smoothing $\lambda=0.1$ with $5$ iterations, (c) quantization $9$ bits, (d) Similarity transformation , (e) Cropping ratio $10.0$.}
    \label{fig:AttackedBimba}
\end{figure}

To evaluate the robustness of the proposed scheme, 3-D meshes have been undergone several attacks. For this purpose, a benchmarking system has been used \cite{benchmark2010}. The robustness of our scheme is tested under several attacks including noise addition, smoothing, quantization, cropping, subdivision and similarity transformations (translation, rotation and uniform scaling).  Fig. \ref{fig:AttackedBimba} shows the model Bimba after several attacks.
To evaluate the robustness to noise addition attack, binary random noise was added to each vertex of 3-D models with four different noise amplitudes : $0.05$\%, $0.10$\%, $0.30$\% and $0.50$\%. According to Table \ref{tab:RobustnessNoiseSmoothEltReord}, it can be seen that the proposed method is robust against noise addition four all the 3-D models.

For evaluating the resistance of the proposed scheme to smoothing attack, the 3-D models have undergone Laplacian smoothing proposed in \cite{taubiny2000geometric} using $5$, $10$, $30$ and $50$ iterations while keeping the deformation factor $ \lambda=0.10$. Table \ref{tab:RobustnessNoiseSmoothEltReord} shows that our method is able to withstand smoothing operation. The robustness of the proposed scheme is evaluated against elements reordoring attack called also file attack. According to Table \ref{tab:RobustnessNoiseSmoothEltReord} the proposed scheme can  resist to element reordoring. Quantization is also applied  to the 3-D models to evaluate the robustness against this attack using  $7$, $8$, $9$, $10$ and $11$ bits. It can be concluded from Table \ref{tab:RobustnessQuantizSimilTrSubdivCropp} that our method shows good robustness  against quantization regardless of the used 3-D mesh. The robustness of our method is evaluated against similarity transformation in which 3-D models have undergone a random rotation,  a random uniform scaling and  a random translation. Table  \ref{tab:RobustnessQuantizSimilTrSubdivCropp} sketches the obtained results in terms of correlation. It can be observed that or method can achieve high robustness against these attacks. Finally, the proposed scheme is tested against subdivision attack including three schemes (loop , midpoint and sqrt3). The obtained results in Table \ref{tab:RobustnessQuantizSimilTrSubdivCropp} in terms of correlation exhibit the high robustness against subdivision. Cropping  is considered to be one of the most damaging attack since it deletes a region from the 3-D mesh and thus the useful information will be lost. It can be observed from Table \ref{tab:RobustnessQuantizSimilTrSubdivCropp} that the proposed method is not enough robust to cropping attacks. In fact, if the deleted surface contains salient points, the extraction process will fail. In the future work, we will search a solution to the issue related to the robustness weakness against this attack.

%Fusion de 3 tables en 1 (Noise,smoothing,elmtReord)

\begin{table}[!h]
\caption{Watermark robustness against additive noise, Laplacian smoothing and elements reordering  measured in terms of correlation.}
\label{tab:RobustnessNoiseSmoothEltReord}
\centering
\resizebox{\textwidth}{!}{%
\begin{tabular}{|c|c|c|c|c|c|c|c|c|c|c|c|}
\hline
 & \multicolumn{4}{c|}{Noise addition} & \multicolumn{4}{c|}{Laplacian Smoothing} & \multicolumn{3}{c|}{Elements reordering} \\ \hline
Parameters & 0.05\% & 0.10\% & 0.30\% & 0.50\% & 5 & 10 & 30 & 50 & Type 1 & Type 2 & Type 3 \\ \hline
Flower & 0.98 & 0.94 & 0.89 & 0.81 & 1.0 & 0.99 & 0.98 & 0.89 & 1.0 & 0.95 & 1.0 \\ \hline
Vase & 1.0 & 0.91 & 0.85 & 0.71 & 1.0 & 0.98 & 0.97 & 0.86 & 1.0 & 0.96 & 1.0 \\ \hline
Cup & 0.96 & 0.93 & 0.83 & 0.78 & 1.0 & 1.0 & 0.90 & 0.87 & 1.0 & 1.0 & 0.98 \\ \hline
Ant & 0.97 & 0.95 & 0.86 & 0.77 & 0.99 & 0.97 & 0.93 & 0.85 & 1.0 & 0.98 & 1.0 \\ \hline
Cat & 0.99 & 0.91 & 0.88 & 0.73 & 1.0 & 0.95 & 0.95 & 0.90 & 1.0 & 1.0 & 1.0 \\ \hline
Bimba & 1.0 & 0.93 & 0.90 & 0.80 & 1.0 & 1.0 & 0.94 & 0.91 & 0.99 & 0.97 & 0.96 \\ \hline
\end{tabular}%
}
\end{table}

\begin{table}[!h]
\caption{Watermark robustness against  quantization, similarity transformations, subdivision and cropping  measured in terms of correlation.}
\label{tab:RobustnessQuantizSimilTrSubdivCropp}
\centering
\resizebox{\textwidth}{!}{%
\begin{tabular}{|c|c|c|c|c|c|c|c|c|c|c|c|c|c|c|}
\hline
 & \multicolumn{5}{c|}{Quantization} & \multicolumn{3}{c|}{Similarity transformation} & \multicolumn{3}{c|}{Subdivision} & \multicolumn{3}{c|}{Cropping} \\ \hline
Parameters & 11-bits & 10-bits & 9-bits & 8-bits & 7-bits & Type 1 & Type 2 & Type3 & Loop iter 1 & Midpoint iter 1 & Sqrt3 iter 1 & 10\% & 30\% & 50\% \\ \hline
Flower & 1.0 & 1.0 & 0.98 & 0.91 & 0.80 & 1.0 & 0.95 & 0.98 & 1.0 & 0.93 & 1.0 & 0.56 & 0.44 & 0.28 \\ \hline
Vase & 1.0 & 1.0 & 0.98 & 0.91 & 0.78 & 0.92 & 0.94 & 0.98 & 0.98 & 0.87 & 0.96 & 0.59 & 0.31 & 0.12 \\ \hline
Cup & 1.0 & 0.99 & 0.97 & 0.93 & 0.83 & 1.0 & 1.0 & 1.0 & 1.0 & 0.84 & 0.94 & 0.67 & 0.37 & 0.16 \\ \hline
Ant & 1.0 & 1.0 & 0.97 & 0.92 & 0.77 & 0.94 & 0.96 & 1.0 & 0.94 & 0.91 & 1.0 & 0.61 & 0.45 & 0.21 \\ \hline
Cat & 1.0 & 0.97 & 1.0 & 0.93 & 0.76 & 1.0 & 1.0 & 0.96 & 0.95 & 0.94 & 1.0 & 0.58 & 0.30 & 0.22 \\ \hline
Bimba & 1.0 & 1.0 & 0.99 & 0.98 & 0.86 & 0.98 & 1.0 & 0.90 & 0.98 & 0.95 & 0.96 & 0.50 & 0.22 & 0.17 \\ \hline
\end{tabular}%
}
\end{table}

\subsection{Comparison with alternative methods}
%\begin{figure}[!h]
%%\centering
%\subfigure[]{
%\includegraphics[height =3.7cm]{images/NoiseComparison.png}
%}
%\quad
%\subfigure[]{
%\includegraphics[height =3.8cm]{images/NoiseComparisonHorse.png}
%}
%\quad
%\subfigure[]{
%\includegraphics[height =3.6cm]{images/SmoothingComparisonVenus.png}}
%\quad
%\subfigure[]{
%\includegraphics[height =3.7cm]{images/SmoothingComparisonHorse.png}}
%\caption {Robustness comparison with Wang's  method \cite{wang2008hierarchical} in terms of correlation against noise attack (a-b) and smoothing (c-d) for Venus and Horse respectively.}
%\label{fig:ComparWangCorr}
%\end{figure}

To further evaluate the performance of the proposed scheme in terms of imperceptibility and robustness we compare it with Cho's \cite{Cho2007}, Wang's et al.  \cite{wang2008hierarchical}, Zhan's et al. \cite{zhan2014blind}, Rolland-Neviere et al. \cite{rolland2014triangle} and Jeongho Son's et al. \cite{son2017perceptual} schemes. We note that for comparison purpose, we have tested the robustness of our method using the 3-D models Bunny, horse and Venus. 
%Je ne vais pas comparer ave cla méthode  nakazawa2010visually car elle ne fournit pas dé résultat et elle utilise juste la méthode de Cho. Nénamoins, on la cite dans les related works.

\begin{table}[!h]
\footnotesize
\centering
{\renewcommand{\arraystretch}{0.25}
%\captionsetup{justification=centering}
\caption{Imperceptibility comparison with Cho's \cite{Cho2007}, Neviere's \cite{rolland2014triangle} and Son's \cite{son2017perceptual}   schemes  measured in terms of MRMS and MSDM for Horse model.}
\label{tab:ImperceptibilityComparison}
%Résultats pour k=750
\begin{tabular}{|c|c|c|c|c|}
\hline
Metric  & \cite{Cho2007} &\cite{rolland2014triangle}&\cite{son2017perceptual} & Our method\tabularnewline
\hline
  MRMS ($10^{-3}$)&$3.17$&$1.48$&$2.90$&$0.53$\tabularnewline
  
\hline
MSDM&$0.3197$&$0.2992$&$0.3197$&$0.2865$\tabularnewline
\hline
 \end{tabular}}\\
\end{table}
Table \ref{tab:ImperceptibilityComparison} exhibits the imperceptibility comparison with schemes in terms of MRMS and MSDM. The obtained results demonstrate the high imperceptibility of the proposed method and show its superiority to the alternative methods.
The proposed method is compared to Cho's \cite{Cho2007} and Zhan's \cite{zhan2014blind} methods in terms of imperceptibility in terms of MRMS as well as robustness in terms of correlation against noise addition, smoothing and quantization using Bunny  and Venus 3-D meshes. Table \ref{tab:NoiseAdditionSmoothQuantizComparison} sketches the robustness comparison in terms of correlation between our method and schemes \cite{Cho2007} \cite{zhan2014blind}. It can be concluded from Table \ref{tab:NoiseAdditionSmoothQuantizComparison} that the proposed method is quite robust to additive noise, smoothing and quantization and outperforms the alternative methods. 
\begin{table}[]
\caption{Robustness comparison with Cho's \cite{Cho2007} and Zhan's \cite{zhan2014blind} schemes  against additive noise, smoothing and quantization in terms of correlation for Bunny and Venus models.}
\label{tab:NoiseAdditionSmoothQuantizComparison}
\centering
\resizebox{\textwidth}{!}{%
\begin{tabular}{|c|c|c|c|c|c|c|c|c|c|c|c|c|c|c|c|c|c|c|}
\hline
Model                                 & \multicolumn{9}{c|}{Bunny}                                                                      & \multicolumn{9}{c|}{Venus}                                                                      \\ \hline
Attack                                & \multicolumn{3}{c|}{Noise addition} & \multicolumn{3}{c|}{Smoothing} & \multicolumn{3}{l|}{Quantization} & \multicolumn{3}{c|}{Noise} & \multicolumn{3}{c|}{Smoothing} & \multicolumn{3}{l|}{Quantization} \\ \hline
Intensity                             & 0.1\%   & 0.3\%   & 0.5\%  & 10       & 30       & 50       & 9         & 8         & 7         & 0.1\%   & 0.3\%   & 0.5\%  & 10       & 30       & 50       & 9         & 8         & 7         \\ \hline
\cite{Cho2007}       & 0.72    & 0.72    & 0.66   & 0.84     & 0.60     & 0.36     & 0.73      & 0.58      & 0.17      & 0.94    & 0.87    & 0.27   & 0.94     & 0.63     & 0.45     & 0.87      & 0.48      & 0.07      \\ \hline
\cite{zhan2014blind} & 1.0     & 0.91    & 0.80   & 0.92     & 0.85     & 0.44     & 1.0       & 0.91      & 0.58      & 0.95    & 0.95    & 0.79   & 0.95     & 0.93     & 0.78     & 1.0       & 0.83      & 0.73      \\ \hline
Proposed method                       & 1.0     & 0.93    & 0.84   & 0.97     & 0.91     & 0.58     & 1.0       & 0.92      & 0.67      & 1.0     & 0.98    & 0.81   & 1.0      & 0.99     & 0.79     & 1.0       & 0.90      & 0.85      \\ \hline
\end{tabular}%
}
\end{table}

\section{Conclusion}
\label{Conclusion}
In this paper, a blind robust 3-D mesh watermarking method based on mesh saliency and QIM quantization for Copyright protection  is proposed. The proposed method  achieves  both high robustness and  imperceptibility. The robustness requirement is achieved  by quantizing the vertex norms using QIM while the imperceptibility achievement is ensured by adjusting the watermarking embedding according to the mesh visual saliency. The experimental results demonstrate that the proposed scheme yields a good tradeoff between the imperceptibility and robustness requirements. Moreover, experimental simulations show that the proposed method outperforms the existing methods against the majority of common attacks. In the future work, we will investigate the issue related to the robustness weakness against cropping attack.
%
%Dans les tableaux il faut remplacer |c| par |l|

 \bibliographystyle{splncs04}
% \bibliography{samplepaper.bib}
%

\end{document}